# Assessing the Case for Africa-Centric AI Safety Evaluations

**Gathoni Ireri[1,2]* Cecil Abungu[1,3]* Jean Cheptumo[1,2]* Sienka Dounia[1,2]* Mark Gitau[1,2] Stephanie Kasaon[4,5] Michael Michie[6] Chinasa T. Okolo[7] Jonathan Shock[2,8]**

[1]ILINA Program [2]African Hub on AI Safety, Peace and Security at the University of Cape Town
[3]University of Cambridge [4]DAIvolve Technologies [5]Action Lab [6]Everse Technology
[7]Technēcultură [8]Dept. of Maths and Applied Maths, University of Cape Town

**Abstract**

Frontier AI systems are being adopted across Africa, yet most AI safety evaluations are designed and validated in Western environments. In this paper, we argue that the portability gap can leave Africa-centric pathways to severe harm untested when frontier AI systems are embedded in materially constrained and interdependent infrastructures. We define severe AI risks as material risks from frontier AI systems that result in critical harm, measured as the grave injury or death of thousands of people or economic loss and damage equivalent to five percent of a country's GDP.

To support AI safety evaluation design, we develop a taxonomy for identifying Africa-centric severe AI risks. The taxonomy links outcome thresholds to process pathways that model risk as the intersection of hazard, vulnerability, and exposure. We distinguish severe risks by amplification and suddenness, where amplification requires that frontier AI be a necessary magnifier of latent danger and suddenness captures harms that materialise rapidly enough to overwhelm ordinary coping and governance capacity.

We then propose threat modelling strategies for African contexts, surveying reference class forecasting, structured expert elicitation, scenario planning, and system theoretic process analysis, and tailoring them to constraints of limited resources, poor connectivity, limited technical expertise, weak state capacity, and conflict. We also examine AI misalignment risk, concluding that Africa is more likely to expose universal failure modes through distributional shift than to generate distinct pathways of misalignment.

Finally, we offer practical guidance for running evaluations under resource constraints, emphasising open and extensible tooling, tiered evaluation pipelines, and sharing methods and findings to broaden evaluation scope.

---

* Lead authors contributed equally. Jonathan Shock's contribution to this paper was supported by the UK government's Foreign, Commonwealth, and Development Office and Canada's International Development Research Centre. The authors thank Victoria Krakovna and researchers at the ILINA program for feedback on an earlier draft of the paper. They also thank Laureen Nyamu for research assistance and Yusra Abdullahi, Faith Gakii and Rugenge wa Nciko for editorial assistance. Corresponding authors noni@ilinaprogram.org and cecil@ilinaprogram.org.

Worryingly, the notion that problems can be solved before they are fully understood still seems widespread.

- Nájera, González-Silva & Alonso,
  Some Lessons for the Future from the Global Malaria Eradication Programme (1955–1969). [1]

# Introduction

The adoption of AI systems is rising globally,[2] with diffusion rates growing across both the Global North and Global South. Even though a gap remains between the two the trend line for both continues to point upward.[3] In Africa specifically, this is evidenced by the increased incorporation of AI in key sectors,[4] the continued creation of national AI policies to guide adoption,[5] and expanding integration of AI systems, particularly in healthcare.[6,7,8]

This widening adoption coincides with advances in frontier AI systems,[9] understood here as highly capable foundation models that could possess dangerous capabilities sufficient to pose severe risks to public safety.[10] Accompanying this rise in performance is a range of emerging risks, including those arising from misuse or loss of control.[11] As frontier AI systems are integrated further into critical sectors, the likelihood of severe AI risks–which we define as material risks from AI systems that result in critical harm–rises in tandem. AI safety evaluations have emerged as one of the principal mechanisms through which severe AI risks are identified before deployment and safeguards implemented before they can lead to harm.[12]

These evaluations work back from anticipated risk pathways to test AI systems, and their findings inform decisions about whether and how to deploy them.[13] Their strength depends on how well their underlying assumptions reflect the conditions of actual deployment.[14] Where those assumptions are drawn from a narrow set of contexts, risks specific to other environments may go undetected. Africa, where frontier AI deployment is accelerating under conditions that differ materially from those in which most evaluations are designed, is one such environment. Although this study focuses on Africa, the underlying problem extends to any region whose conditions are not adequately reflected in evaluation design.

Recent scholarship has advanced our understanding of both the categories of severe AI risk[15] and the methodologies available to evaluate them,[16] while a parallel body of work has exposed the limits of current evaluations, notably their tendency to test algorithmic outputs in isolation from the sociotechnical systems in which AI operates[17] and to assume that results obtained in one context transfer reliably to others.[18] A growing literature has also drawn attention to the need for AI safety efforts that account for the conditions of the Global South,[19] where deployment environments diverge considerably from the high-resource settings in which most evaluations are designed. What remains absent, however, is a framework that connects these concerns: one that identifies which severe AI risks are most salient in African contexts, models the pathways through which they could arise, and informs evaluation design accordingly. This paper attempts to fill that gap.

The paper is organised in seven main parts. The first reviews severe AI risks and considers the limitations of current evaluations. The second develops criteria for identifying severe AI risks for evaluation. The third suggests threat-modelling strategies applicable to African contexts. The fourth considers whether misalignment risks vary across contexts. The fifth suggests tools and frameworks for researchers interested in conducting evaluations. The sixth sets out recommendations for key actors responsible for developing, assessing, and governing AI systems. The seventh concludes with a summary of our main contributions.

# 1. Severe AI risks and the limits of current evaluations

## 1.1 High-level definition of severe AI risk

Severe AI risks can be defined as material risks from frontier AI systems that result in critical harm. In this definition, frontier AI systems are highly capable foundation models that may possess dangerous capabilities sufficient to pose severe risks to public safety,[20] risk refers to the probability of an outcome worse than the status quo, and critical harm refers to outcomes resulting in grave bodily injury, death, or significant economic loss or damage. This framing is informed by scholarship on catastrophic risks, legal definitions, and industry practice.

In the catastrophic risk literature, catastrophes have been defined as sudden and dramatic incidents that disrupt the normal course of events and profoundly shift the way in which society is organised,[21] typically resulting in loss of life and extensive damage.[22]

These conceptions of catastrophe provide a reference point for how severe AI risks are framed in practice. In safety and preparedness frameworks, which outline how risks are anticipated and managed as model capabilities advance, leading frontier AI developers reference severe and catastrophic risks.[23] This approach is consistent with the Frontier AI Safety Commitments[24] made at the AI Seoul Summit in 2024, which aimed to garner voluntary commitments from AI developers to develop and deploy safe and trustworthy AI systems[25] by publishing a safety framework focused on severe risks.

For instance, Anthropic's Responsible Scaling Policy focuses on preventing catastrophic risks, defined as large-scale devastation (thousands of deaths or hundreds of billions of dollars in damage) directly caused by an AI model that would not have occurred without it.[26] OpenAI's Preparedness Framework[27] states that its safety commitments track and prepare for capabilities that create new risks of severe harm, which they define as the death of or grave injury to thousands of people or economic damage amounting to hundreds of billions of dollars. Similarly, Google DeepMind's Frontier Safety Framework is designed to address severe risks that may arise from powerful capabilities of future foundation models.[28] Its approach considers severe risk with reference to the capability level at which misuse,[29] misalignment,[30] and machine learning research and development (ML R&D)[31] risks could prove severe.

Legal and policy frameworks have also referenced severe risks and harm from AI systems. In September 2025, the California Senate passed the Transparency in Frontier Artificial Intelligence Act (SB-53),[32] which defines catastrophic risk as a foreseeable and material risk that a frontier developer's development, storage, use, or deployment of a frontier model will materially contribute to the death or serious injury of more than fifty people, or more than one billion dollars in damage to or loss of property. Similarly, the New York Senate recently enacted the Responsible AI Safety and Education (RAISE) Act,[33] which defines a safety incident as a known incidence of critical harm, defined as the death or serious injury of one hundred or more people, or at least one billion dollars in damages to rights in money or property.

Informed by these perspectives, we define severe AI risks as material risks from frontier AI systems that result in critical harm, measured by the grave injury or death of thousands of people, or economic loss and damage equivalent to 5% of a country's GDP. We discuss the defining characteristics of severe AI risks in greater detail in part 2.

## 1.2 Perspectives taken by industry and policy institutions

The ways in which AI systems may contribute to catastrophic risk have been classified into four categories based on their mechanisms of emergence.[34] These categories are malicious use, AI race dynamics, organisational risks, and rogue AI.[35] Malicious use involves the intentional deployment of AI systems to cause harm, including enabling large-scale biological, cyber, or information attacks. AI race risks arise from competitive pressures that incentivise rapid deployment at the expense of safety, potentially leading to destabilising arms races. Organisational risks arise from accidental failures or governance breakdowns within AI-developing institutions. Finally, rogue AI risks concern harmful behaviour originating within the AI systems themselves, such as goal misalignment, deception, or loss of control as capabilities increase. Although this framework is analytically useful, in practice, AI developers often prioritise certain risk categories over others in their safety evaluations.

For example, Anthropic's Responsible Scaling Policy[36] identifies and evaluates two severe AI risks they consider the 'most pressing catastrophic risks': chemical, biological, radiological and nuclear (CBRN), and Autonomous AI Research and Development (AI R&D). OpenAI's Preparedness Framework[37] approaches risk assessment and monitoring by dividing risks into tracked and research categories. Their tracked categories are biological and chemical risks, cybersecurity, and AI self-improvement, while their research categories are long-range autonomy, sandbagging, autonomous replication and adaptation, undermining safeguards, and nuclear and radiological threats. Google's Frontier Safety Framework[38] specifies three severe AI risk categories: misuse (CBRN, cyber, and harmful manipulation), machine learning research and development, and misalignment risk. METR's analysis of frontier AI safety policies notes that biological threats, cyber capabilities, and automated AI R&D consistently appear among the severe risk categories identified by developers.[39]

Beyond industry-led frameworks, public and independent institutions have articulated similar concerns about severe AI risks, reinforcing the view that these risks are not merely speculative. In its research agenda, the UK AI Security Institute (UK AISI) states its commitment to delivering rigorous research for the most serious emerging risks from AI, including cyber risks, chemical and biological risk, criminal misuse, and risks from autonomous systems.[40] While some organisations address "severe risks" broadly, others explicitly prioritise the most extreme cases. For example, Apollo Research has articulated its mission as reducing "catastrophic risks posed by advanced AI"[41] and more recently, as understanding and addressing "extreme risks from frontier AI systems."[42] Similarly, METR's mission is to develop scientific methods to assess catastrophic risks stemming from autonomous capabilities of AI systems and enable good decision-making about their development.[43]

We highlight these policies and institutional approaches for two reasons. First, they demonstrate a broad recognition that frontier AI systems may pose critical or catastrophic harms. Second, they provide an overview of how leading developers and research organisations are currently structuring, categorising, and evaluating these risks. Accordingly, there has been a concerted effort to evaluate frontier AI systems as a means of identifying and mitigating severe AI risks. Central to this effort are safety evaluations.

## 1.3 AI safety evaluations and severe AI risk governance

AI safety evaluations can be defined as comprehensive tests that work back from identified risk pathways and attempt to pinpoint failures in AI systems that could be exploited to cause harm.[44] They are intended to assess how safe a given AI model is and can be classified in several ways.

Ji et al. distinguish between model safety evaluations, which evaluate the outputs of models alone, and contextual safety evaluations, which evaluate how models impact real-world outcomes such as user behaviour and decision-making.[45] Evaluations can also be classified by properties such as capability, propensity, and control. Capability evaluations measure what a model is able to do and how those abilities inform its potential risk profile,[46] testing for abilities such as deception, persuasion and manipulation, situational awareness, and political strategy.[47] Propensity evaluations, also known as alignment evaluations, assess what a model tends to do by default, often by examining model behaviour when given choices between different actions.[48] Control evaluations assess whether safety protocols remain effective when models intentionally try to override them.[49]

These evaluations serve several functions, including improving internal processes and governance, providing assurance to external stakeholders that AI systems are trustworthy, and validating claims of trustworthiness.[50] Beyond these functions, they have a significant bearing on the AI governance approaches taken for these systems, both internally (by AI developers) and externally (by policymakers).

With respect to policymakers, evaluations ostensibly provide the evidence base for informed policy decisions[51] whereas, with respect to AI developers, risk analyses from evaluations inform decisions such as whether to release a model, how to release it, and what uses should be permitted. These assessments therefore influence regulatory and deployment decisions. Consequently, if an evaluation system is not robust, the decisions relying on it may be misguided.

This importance of AI safety evaluations is reflected in recent legislation. The EU AI Act sets out requirements for high-risk AI systems[52] including a mandatory risk management system,[53] described as a continuous iterative process that runs throughout the entire lifecycle of a high-risk AI system comprising, among others, the evaluation of risks that may emerge and of other risks possibly arising based on data analysis from post-market monitoring systems.[54] The New York RAISE Act requires frontier AI developers to have a safety and security protocol describing in detail the testing procedure to evaluate if the frontier model poses an unreasonable risk of critical harm, making this a prerequisite for deployment.[55]

Similarly, California's SB-53 requires frontier developers to publish a frontier AI framework that, among other things, defines and assesses thresholds used to identify and assess whether the frontier model has capabilities that could pose catastrophic risk.[56] Likewise, the NIST AI Risk

Management Framework emphasises evaluation, testing, and continuous monitoring as core components of AI risk management across the lifecycle of development and deployment.[57] Across both legislative requirements and voluntary risk management frameworks, evaluation is treated as a central mechanism for managing frontier AI risk. The strength of this mechanism, however, depends on the assumptions underlying evaluation design.

## 1.4 The portability challenge in AI safety evaluations

Evaluations are often designed within specific contexts and assumed to apply beyond them, yet they have been faulted for lacking external validity by failing to generalise beyond their testing context.[58] This limitation has been described as the "framing trap", where evaluations test algorithmic outputs without modelling the sociotechnical system in which the AI operates, including local infrastructure and societal norms.[59] The framing trap is compounded by the "portability trap", where solutions evaluated in one social context may be harmful or misleading when applied elsewhere.[60]

Similarly, there is concern about "unknown unknowns".[61] These are potential risks that current evaluation protocols might miss, based on the observation that current evaluation methods are limited by our ability to anticipate what we need to evaluate.[62] Although Grey and Segerie discuss this in the context of AI systems potentially developing unanticipated capabilities, contextual peculiarities—such as language and form (for example poetic form versus continuous prose)—may also warrant evaluation. Likewise, combinatorial complexity can mean that one or more unaccounted-for factors combine with accounted-for factors to create an "unknown unknown" scenario.[63] Grey and Segerie pose the question: "if we can only explore 0.001% of the interaction space, how can we be confident we have identified the most dangerous emergent patterns?"

Our response to this question is that we cannot be confident, which underscores the need for evaluations that account for a wider range of interaction scenarios. At a minimum, this may result in a more accurate estimation of our confidence in current evaluations; at best, it may inform training and mitigation strategies to improve safety across differing deployment contexts. We do not claim the interaction space can ever be completely covered, but a greater portion of it should be accounted for through evaluations. This is especially true where deployment conditions differ significantly from those assumed in evaluation design.

## 1.5 Severe AI risks and Africa's deployment context

AI adoption across Africa is likely to accelerate in the coming years as part of a broader global expansion in AI uptake. In August 2024, the African Union (AU) launched a Continental Artificial Intelligence Strategy.[64] In the strategy, the AU proposes to accelerate the adoption of AI in the public sector and in sectors with high social and economic value, including agriculture, education, and health.[65] At the national level, several African countries have developed AI strategies, in which they express an intention to harness AI to improve the delivery of services in sectors such as healthcare, transportation, and education.[66] While such strategies do not guarantee widespread or rapid implementation, they signal a policy direction that could significantly expand the contexts in which AI systems operate and are relied upon.

These policy commitments coincide with growing external investment involving increasingly capable AI systems, including frontier AI. For example, in January 2026 the Gates Foundation and OpenAI announced a fifty million dollar ($50 million) initiative to deploy AI tools across 1,000 primary healthcare clinics in Rwanda and other African countries, with the stated aim of alleviating chronic health-worker shortages and supporting clinical decision-making.[67] This illustrates how AI is intended as a capacity-augmenting tool in resource-constrained settings.

However, the outcomes of frontier AI deployment in such contexts will depend, in large part, on the conditions of the systems into which these technologies are introduced.[68] This is particularly salient in relation to critical infrastructure. By critical infrastructure, we refer to systems, facilities, and assets essential to the functioning of society and the economy.[69] In this analysis, and with respect to critical infrastructure, "systems" denotes the interconnected assets, networks, and facilities that collectively enable critical functions.[70] A system is considered weak when its ability to deliver essential services is constrained by factors such as unreliable resources including inconsistent power supply, inadequate equipment, insufficient personnel, or limited institutional capacity.[71]

Prior work has shown that prevailing definitions of critical infrastructure often adopt an overly narrow, technical, or sector-based view, omitting human factors, external dependencies, and broader political and cultural environments that shape how infrastructures function in practice.[72] These omissions can obscure sources of vulnerability and risk, particularly in contexts where systems operate under constrained conditions or strong interdependencies.[73] Building on this perspective, we distinguish between two forms of criticality. Inherent criticality refers to infrastructures whose disruption would directly result in societal, economic, or safety consequences such as electricity supply, water systems or emergency healthcare services.[74]

By contrast, external criticality arises from interdependencies between infrastructures, where failure in one system propagates disruption across others.

On this basis, many African critical infrastructure systems combine high inherent criticality and high external criticality, while simultaneously operating under more constrained conditions. For example, only 40% of health facilities in sub-Saharan Africa have access to reliable electricity,[75] while the region has only 1.55 physicians, nurses and midwives per 1,000 people, far below the World Health Organisation threshold of 4.45 needed for universal health coverage.[76] Beyond healthcare, fewer than half of Africans receive a supply of electricity that works "all or most of the time," with lower rates of reliable electricity among rural households, and a general dissatisfaction with government electricity provision and national grid performance.[77]

These infrastructure constraints do not exist in isolation. Research demonstrates that infrastructure interdependence means failures affecting a single infrastructure can cascade across systems.[78] According to Yan *et al.*, urban infrastructures are interconnected across cyber, functional, physical, spatial, logical, and organisational dimensions.[79] Taken together, constrained baseline infrastructure coupled with high interdependency means that many African critical infrastructure systems operate under conditions that differ materially from the relatively stable, high-capacity environments in which frontier AI systems are typically developed and evaluated.[80] This difference creates an infrastructure gap between deployment environments and evaluation contexts.

This gap has important implications for AI safety evaluations. When frontier AI systems are deployed in environments characterised by high criticality, errors, misrepresentations, or inappropriate recommendations can propagate across interconnected systems and amplify risks. In such cases, safety failures may not arise because AI systems malfunction in a narrow technical sense, but because the conditions under which their safety was assessed fail to capture the interdependencies, human factors, and contextual constraints present in deployment environments. In this context, the state of Africa's critical infrastructure directly shapes how AI adoption will unfold on the continent.[81]

The current scope of AI safety evaluation represents decisions about what to prioritise in testing. Deliberately including African contexts, low-resource languages, and infrastructure scenarios typical of the Global South would substantially increase the tested interaction space beyond its current fraction. This expansion matters because safety determinations shape deployment decisions, and those determinations should reflect the full range of contexts where these systems will operate. The aim is not to achieve perfect coverage, which combinatorial complexity makes impossible, but a broader and more representative evaluation scope. We consider African contexts to be underrepresented in current evaluation designs.

Taking healthcare as an illustrative case, both the potential benefits and risks of deploying frontier models such as ChatGPT in low- and middle-income country health systems have been discussed.[82] Africa experiences more than 160 public health emergencies each year, most of which are infectious diseases requiring coordinated responses across multiple components of the health system.[83] Yet, only around 1% of global health data originates from African countries, while much of AI research and model training draws predominantly on Western and Chinese datasets.[84] As a result, AI systems deemed "safe" based on evaluations conducted primarily in high-capacity clinical and infrastructural settings may encounter contexts characterised by different patterns of disease prevalence and public health response structures that were not incorporated into evaluation design.[85]

Clinical AI research cautions that applying models trained on homogenous datasets to socio-demographically distinct populations, without external validation, poses a considerable risk.[86] Predictions may be inaccurate when underlying data distributions differ, and under-represented populations may therefore be disadvantaged when excluded from the datasets used to build and evaluate these systems.[87] It may be argued that reliance on Western data is not unprecedented, given that medical education in many non-Western contexts, including Africa, draws on globally produced research. However, clinical practice is characterised by iterative, context-specific learning and adaptation to local settings,[88] whereas AI systems may not reliably recalibrate to local epidemiological conditions without deliberate training and evaluation.[89] Further, African health systems often lack the data infrastructure, computational capacity, and governance frameworks needed to support reliable AI integration, including high-quality health data, and robust internet connectivity.[90]

Accordingly, until evaluation scope expands, determinations of what is "safe for deployment" will continue to reflect evaluation performed in high-resource contexts that may not accurately reflect differing deployment environments, including African contexts. While similar structural conditions may exist elsewhere in the Global South, the analysis here focuses on Africa given the policy momentum and frontier AI deployment initiatives currently underway on the continent. Without meaningful African participation in evaluation design and execution, frontier AI systems risk being deployed on the basis of incomplete safety assessments that overlook severe AI risk pathways specific to African settings. Therefore, incorporating African contexts into evaluation frameworks is necessary to reduce the likelihood of failures arising from deployment contexts not adequately represented in current evaluation designs. This requires, first, a means of identifying which severe AI risks are most relevant to these contexts.

# 2. Criteria for identifying Africa-centric severe AI risks

Determining which severe AI risks should be prioritised requires a framework that accounts for both the emergence of a risk and the threshold at which it would constitute critical harm. The framework developed here sets out outcome measures of harm alongside the process pathways by which a risk arises. Outcomes denote critical harms and are operationalised as material risks from frontier AI systems that could result in the sudden death or grave injury of thousands of people, or economic loss and damage equivalent to 5% of a country's GDP. Process pathways refer to the complex intersection of vulnerability, hazard, and exposure that, in combination, can trigger the materialisation of risk.[91]

The framework additionally sets out two distinctive features of a severe AI risk, intended to distinguish this category from other kinds of risk: time and amplification. Time, as a function of suddenness, accounts for harms that occur so suddenly as to seem discontinuous with the flow of previous events.[92] Amplification[93] captures the contribution of frontier AI to the materialisation of risk, implicitly assuming that AI serves as a magnifier of latent risk. Under this framing, the tag 'severe AI risk' applies only in cases where critical harm would not have occurred but for the AI system. The rationale behind setting such a high causal bar is the practical consideration that resources and interventions should be concentrated at the point where risk is greatest.

We explain how we arrived at these factors in the sections that follow.

## 2.1 Outcome measures of critical harm

The definition of outcomes hinges on harm as a function of grave bodily injury, death, or economic loss and damage caused. These harms are categorised as 'critical' when they exceed a set threshold. Drawing from the global catastrophic risk literature, Sundaram & Mani identify three key definitions of outcome harms: events causing 10 million deaths or $10 trillion in economic loss, the loss of 10% of the global population, and the loss of 25% of the global population.[94] In AI safety policy, similar thinking plays out at different magnitudes. As detailed in section 1.1, Anthropic and OpenAI set thresholds at thousands of deaths or hundreds of billions in damages, while California's SB-53 (50 deaths, one billion dollars) and New York's RAISE Act (100 deaths, one billion dollars) establish a lower bar.

Overall, definitions of global catastrophic risks vary, particularly in relation to thresholds of harm. This has led to a wide range of disparate thresholds that invite scrutiny, for instance, whether 10% population loss is a sound anchor for catastrophic loss of life, or what should mark the lower limit for economic loss and damage. What proves most useful, however, is the measurability aspect: in most cases, lives lost and economic damage serve as proxies for direct harm at scale.

To calibrate the threshold for critical harm as it applies to injury and death, major human-caused catastrophes from the past decade in select African countries provide a useful reference point. In the Tigray War, attacks in Axum on 28–29 November 2020 are estimated to have resulted in the deaths of hundreds of civilians in approximately 24 hours.[95] In Nigeria, Boko Haram fighters reportedly killed between 150 and 2,000 people in the town of Baga within five days.[96] Over longer timeframes, Côte d'Ivoire's 2010 post-election crisis is believed to have led to at least 3,000 deaths across five months,[97] while Burundi's 2015 political crisis resulted in an estimated 700 to 1,000 deaths within a year.[98]

These cases demonstrate that critical harm is dependent on concentration, which can be understood as a function of both scale and time. A tiered definition is therefore adopted: the grave injury or death of thousands over months serves as the primary threshold, while the grave injury or death of hundreds within the span of weeks equally qualifies as critical harm. Since AI-driven harms are more likely to be distributed over months than concentrated in single incidents, the first threshold serves as the primary definition, recognising that shorter-duration mass casualty events also constitute catastrophes by historical standards.

The adopted bodily harm and loss of life thresholds largely mirror those adopted by Anthropic and OpenAI but differ from the legislative thresholds set out by California and New York. For economic harm, a different standard is adopted, set at 5% of an individual country's GDP in economic loss and damage. This reflects the view that current economic thresholds are set so high as to be irremediable, thereby exceeding critical harm and entering the realm of existential catastrophe.

Consider that the current threshold set by frontier AI companies (hundreds of billions of dollars in damage) would, in almost all cases, be impossible for most African countries to recover from. If a catastrophe were to strike that caused $200 billion in economic damage, how would that affect Kenya differently than it would the UK? For Kenya, with a GDP of approximately $130 billion,[99] this represents 150% of annual economic output, whose loss would be an existential crisis that would exhaust the country's $10 billion in foreign reserves,[100] risk currency collapse, and likely require a decades-long international bailout. For the UK, with a GDP of approximately $3.6 trillion[101] and $175 billion in reserves,[102] the same $200 billion disaster represents 6% of annual output. After exhausting all reserves, the remaining $25 billion shortfall could be financed through borrowing in international credit markets, where insurance would cover a

significant portion, likely allowing recovery within a significantly shorter period without presenting an existential threat to its institutions or currency.

The proposed threshold would be applicable even for the country with the lowest GDP in Africa, São Tomé and Príncipe, which stands at approximately $822 million.[103] Using this heuristic, severe AI risk evaluations for São Tomé and Príncipe would be checked against $40 million in economic loss and damage. While the precise level is debatable, 5% of country GDP has been chosen as it is low enough that recovery is likely, but not so low as to capture a broad range of risks that would no longer warrant the distinction of "severe". The aim is to allow the identification and evaluation of risks at the extremes.

Precedent set by New York and California demonstrates that thresholds can be revised depending on context, and the threshold proposed here may be adjusted iteratively if it becomes clear that extreme cases are being missed. Both jurisdictions have diverged from the hundreds-of-billions thresholds proposed by AI companies, adopting instead their own standards set at 'greater than $1 billion in economic damage.[104]

Accordingly, the recommended definition of critical harm for Africa-centric severe AI risks is the grave injury or death of thousands of people, or 5% of annual GDP in economic loss or damage. Researchers and evaluators are encouraged to adopt this definition when designing evaluations.

## 2.2 Process pathways of latent risk

Process pathways refer to the complex intersection of factors that, in combination, can trigger the emergence of severe risks.[105] This framing assumes that the materialisation of risk is context-dependent, as relevant factors vary across locations and societies. An appreciation for process pathways therefore allows for more accurate mapping of risks and facilitates the identification of appropriate mitigations. The central factors in this component of the framework are hazard, vulnerability, and exposure.[106] A hazard is the source of danger; a vulnerability is the weakness within a system that allows the danger to materialise; and exposure is the surface at which the hazard interacts with the vulnerability.[107]

A useful illustration of how these factors can intersect to trigger the emergence of severe risk is the case of malaria in Africa. In 2021, malaria ranked fourth among causes of death in Africa but did not feature in the top five in other continents.[108,109] This was not always the case, as malaria was once common across half the world before being eliminated in many regions.[110] One effort contributing to this elimination was the Global Malaria Eradication Programme of 1955–1966, led by the World Health Organisation.[111] The campaign largely adopted an eradication-first strategy and applied it globally with minimal adaptation, even to areas that would have benefited from a control approach, that is, areas where it may have been more effective to reduce disease incidence rather than attempt elimination through sustained spraying. As a result, while the program succeeded in many regions, it failed in others, including Africa.

Applying the process pathways framework to this case, the hazard would be the same, but the vulnerability and exposure surface would differ. In Africa, an inherent vulnerability during the campaign was a weak public health system, which was needed to address existing cases even as funding was channelled toward eradication. Similarly, the exposure surface was large: many areas were affected by malaria while simultaneously lacking robust public health infrastructure. A failure to account for these intersecting factors likely led to an overemphasis on the hazard itself when the harm arose from the combination of all three. As Nájera et al. observed, "extrapolation of early local experiences, although successful, represented a very limited variety of epidemiological situations…it was obvious from the start that nobody knew how to deal with the problems of tropical Africa".[112]

In the context of severe AI risks, process pathways function to conceptualise these contextual differences and can additionally serve to build out credible threat models, as discussed in section 3.3. For example, the source of danger (hazard) might be conceptualised as the malicious use of a model. Having identified the hazard, the next step would be to consider what weaknesses in the system would contribute to this risk (vulnerability) and finally, the extent to which people and systems are exposed to harm given these weaknesses (exposure).

As vulnerability, hazard, and exposure shift depending on social-structural factors, the ways in which they could interact are nearly inexhaustible. Researchers and evaluators are therefore encouraged to consult the threat modelling methodologies we set out later in this paper.

## 2.3 Distinctive features of severe AI risk

While outcome and process pathways are central to the proposed taxonomy of severe AI risks (critical harm is checked against a set threshold and presumed to be a function of context), two auxiliary characteristics help distinguish severe AI risks from other categories of AI risk: amplification and time.

Borrowing from Arnescheidt et al., amplification is a mechanism by which a hazard may result in harm greater than its direct impact alone.[113] In this framework, the amplifier is exclusively frontier AI. To establish that a severe AI risk is plausible, that is, a risk capable of producing harm that crosses our thresholds, it is necessary to demonstrate that the threshold could only be crossed due to the presence or use of frontier AI. This draws from the 'but-for' test, a causal standard often used to determine liability, which asks whether harm is more likely than not attributable to a significant factor. Given that frontier AI systems are likely to contribute to a range of risks, it is necessary to distinguish instances where they form the core of the risk and, in their absence, harm would not materialise. This distinction is informed by the practical consideration that correctly identifying the main source of risk is likely to have the greatest effect in forestalling future harm.

The second distinctive feature is time, characterised here as "suddenness", harm arising in a way that seems discontinuous with preceding events.[114] The severity of AI risks is considered a function of the concentration of harm, determined by both scale and time. The underlying assumption is that harm spread over a long period may be easier to manage and recover from than harm that materialises rapidly.

This framing is similarly informed by practical considerations. Without a temporal dimension, certain harms would meet our thresholds but over a much longer period, for example, 1,000 deaths over a decade. While such harm could satisfy our other criteria (would not have occurred without frontier AI; arose through the combination of hazard, vulnerability, and exposure), it might be better understood as a societal impact of AI systems rather than a severe AI risk. The methodology for addressing long-term harms differs from that for assessing short-horizon harms and falls outside the scope of this taxonomy. This taxonomy is largely conceived to account for risks that could cause harm at scale within a relatively short period.

Overall, the framework for severe AI risk consists of three components. The first is outcomes, measured as a function of the scale of harm. The second is process pathways, shaped by the interaction of hazard, vulnerability, and exposure. The third is a set of distinctive features: amplification, where frontier AI serves as the mechanism by which harm exceeds its direct impact, and time as a function of suddenness, indicating discontinuity with the ordinary flow of events.

A framework is favoured over a definitive list because it allows researchers to evaluate any risk systematically, testing whether and how its features could lead to critical harm. One approach is to work backwards from known harms, mapping out the process pathways through which they arose and then determining whether and where frontier AI could significantly amplify existing risk. Where a plausible case exists, threat modelling and targeted evaluations follow. This process is detailed in the diagram below.

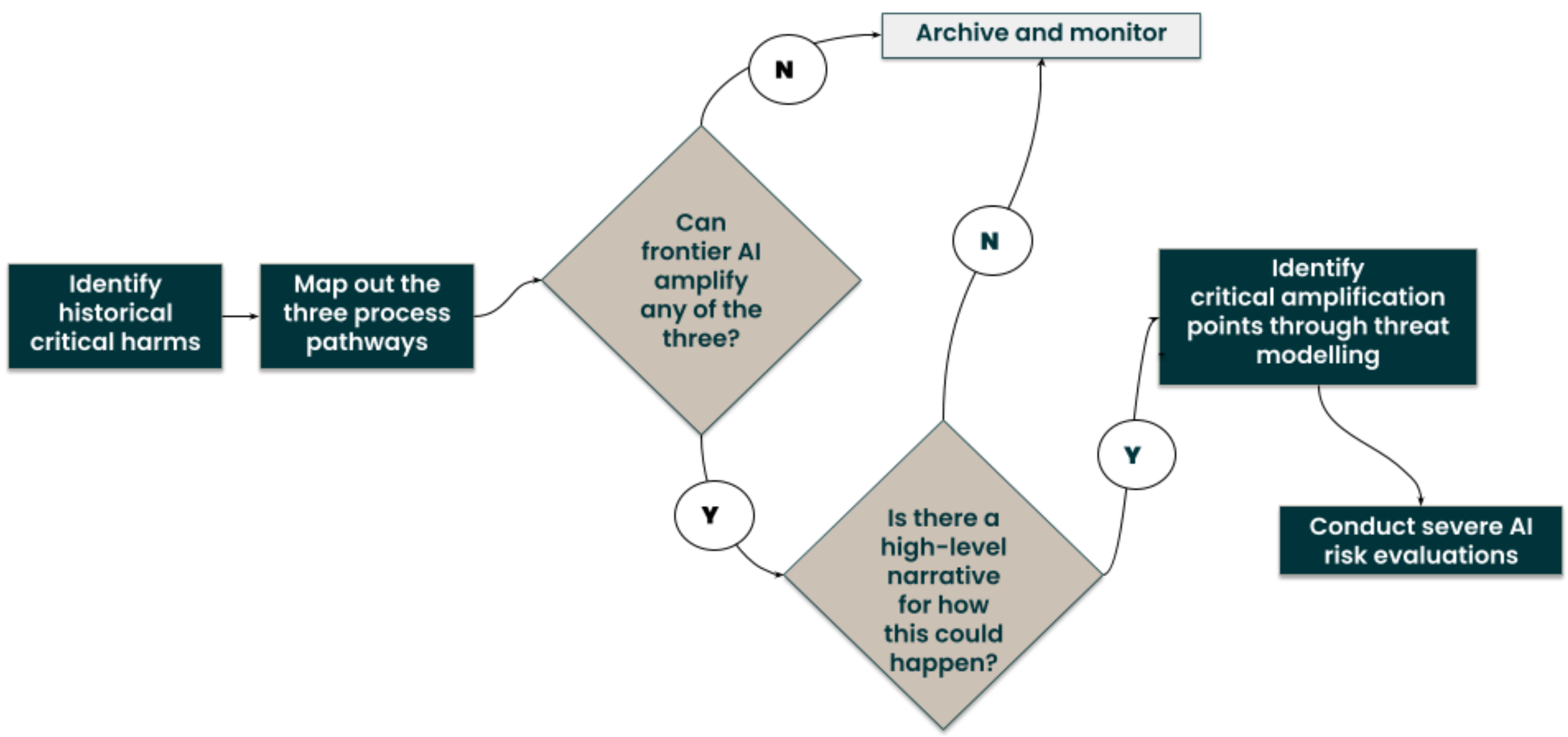

While the process sketched above has the advantage of being grounded in historical precedent, it is limited in its ability to account for genuinely novel risks. Since this framework focuses primarily on amplification, the capacity of frontier AI to magnify existing risks, it does not attempt to illustrate a process for evaluating entirely new categories of harm. Other researchers and evaluators are encouraged to build on this framework to address that gap.

Where the framework does contribute is in informing the design of structured threat models: process pathways can help identify possible critical amplification points, while the proposed thresholds provide a countercheck for conceptualising harm.

# 3. Threat modelling severe AI risks in African contexts

## 3.1 An introduction to threat modelling

Safety evaluations rely heavily on threat models,[115] which provide a structured way of identifying potential risks to a system and specifying mechanisms for mitigation. As a practice, threat modelling consolidated during the Second World War across institutions within and adjacent to the United States and United Kingdom militaries, where early versions were used to anticipate possible actions by Axis powers.[116] Its practical value makes it adaptable across a wide range of domains, with the underlying question remaining largely unchanged: what would an intelligent and determined adversary do?

Three broad approaches are commonly used to answer this question. Asset-centric models focus on the resources or assets that need protection. Threat-centric models emphasise the characteristics, capabilities, and motivations of potential attackers. System-centric models examine how interactions between system components create vulnerabilities. While these approaches can be applied separately, they're typically used together, with contemporary practice favouring a combined rather than isolated analysis.[117]

These approaches are relevant in the context of advanced AI systems. As outlined in the proposed framework, when advanced AI systems act as amplifiers of existing risk, they can substantially alter risk pathways by increasing the capabilities of hazards, exploiting existing vulnerabilities more efficiently, or expanding the exposure surface. In addition, AI systems may themselves present as new risk vectors, a possibility examined in the discussion of misalignment (part 4). Because these risk pathways are highly context-specific, they offer a broad foundation for developing Africa-centric threat models.

How these models are constructed, however, depends on the threat-modelling techniques employed. The choice of technique shapes both the quality and utility of severe AI risk evaluations. The sections that follow survey standard approaches in severe AI risk assessment before examining their applicability to African contexts and identifying which hold the greatest promise under some common African contexts. Throughout this analysis, we focus on severe AI risks - those crossing our thresholds for critical harm and characterised by suddenness rather than gradual degradation.

## 3.2 Standard threat modelling techniques

### 3.2.1 Reference class forecasting

Reference Class Forecasting (RCF) predicts outcomes by examining what typically happened in similar past situations, providing an outside view based on historical patterns rather than focusing on AI's unique features.[118] This approach helps counter optimism bias—the tendency to expect best-case outcomes while ignoring failure rates.[119] It works by selecting an appropriate reference class, examining typical outcomes in that class, and then adjusting for what is genuinely novel about AI.[120]

Tetlock's research demonstrates that reference classes improve forecasting accuracy, although the method struggles when data is scarce or when systems are genuinely novel.[121] Some scholars address this limitation by scaling data from analogous projects to approximate missing evidence.[122] Adapting RCF for African contexts means creating context-specific reference classes that reflect regional risk patterns. We argue that relevant reference classes can be drawn from historical events where latent risks were suddenly amplified, resulting in mass casualties. In African contexts, this could include episodes of mass violence where technological tools caused inflammatory content to spread faster than institutions could respond, or cascading infrastructure failures that caused sudden loss of life at scale.

### 3.2.2 Structured expert elicitation

Structured Expert Elicitation (SEE) enables predictions in the absence of historical data by systematically gathering and synthesising expert judgement. The method uses structured steps that allow experts to quantify their uncertainty about risks, and then weights their views based on how accurately they performed on test questions with known answers.[123] The strength of SEE lies in converting uncertainty into clear probabilities, though its weakness is that results may be biased depending on who qualifies as an "expert".

Adapting SEE for African contexts requires including local experts rather than relying solely on Western technical experts who may not adequately understand the local contexts. One promising approach is a hybrid model that separates technical and contextual risk estimation.[124] Using this approach, researchers can rely on Western AI experts for estimates of how a model's capabilities could change over time and rely on African biosecurity specialists, public health officials, and conflict analysts for assessments about how AI capabilities might amplify latent risks in their domains.

The latter could estimate, for instance, the number and sophistication of potential adversaries who might exploit AI-enabled biological or chemical information given local conditions, the vulnerability of populations to AI-facilitated mass manipulation, and the extent of the absence of structures that could prevent rapid escalation to mass casualty events. Combining these perspectives produces risk estimates grounded in local realities rather than abstract theory. Of course, many African countries may lack deep pools of trained domain experts. In such cases, well-respected local research centres offer potential sources of expertise. We address the implications of this constraint for method selection in section 3.3.3.

### 3.2.3 Scenario planning

Scenario planning is a method where one prepares for risk by imagining several plausible futures instead of attempting to predict a single outcome.[125] The method is particularly useful when problems are complex and uncertain, as with AI risks, because it allows policymakers to test whether proposed responses would remain viable under different conditions. Its strength lies in helping policymakers anticipate risks and develop resilience strategies,[126] while its major weakness is the difficulty of mapping complex scenarios of gradual degradation. The method can also be challenging to implement given its qualitative nature.[127]

In African contexts, scenario planning can focus on infrastructural fragility and governance gaps while using participatory workshops with local communities to ensure that the mapped scenarios reflect the lived realities of affected populations. This aligns with emerging threat modelling best practices that advocate for the inclusion of diverse stakeholders.[128] The method can be complemented by Hybrid Threat Modelling (HTM), which combines multiple analytical techniques while accounting for the specific cultural context, processes, and feedback loops of a given system.[129]

### 3.2.4 System theoretic process analysis

System-Theoretic Process Analysis (STPA) is a method that treats safety as a control problem rather than a reliability issue, examining how accidents occur when control structures fail, instead of focusing on individual component failures.[130] In the context of AI, the process involves defining hazards, modelling the control loops between AI, humans, sensors, and actuators, identifying unsafe control actions, and analysing why such actions might occur. The method's strength lies in identifying risks that emerge from interactions between system components while incorporating human, management, and regulatory factors.[131] However, it is resource-intensive and relies on the analyst's knowledge of the system.[132]

In African contexts, STPA proves particularly useful because it can incorporate environmental instability as a key variable. This includes factors that create vulnerability to sudden, catastrophic failures, for example internet bandwidth issues that prevent humans in the loop from effectively controlling AI agents, intermittent electricity that could disable containment or early warning systems, limited sensor networks that delay detection of biological or chemical releases until casualties mount, and oversight gaps that allow dangerous capabilities to reach adversaries without intervention. The method demonstrates that relying on human intervention is often unsafe in resource-limited settings[133] and instead recommends fail-safe designs that do not assume consistent human oversight.

## 3.3 Proposed threat modelling techniques based on common African contexts

In this section, we match the above techniques to the specific contexts most likely to shape decision-makers' priorities. We focus on contexts that are common across many African countries and that, while not exclusive to the continent, are particularly acute in their scale.

### 3.3.1 Limited organisational resources[134]

When resources are scarce, RCF and qualitative scenario planning should form the methodological core because both can be implemented at minimal cost without scientific rigor. RCF works by identifying analogous historical cases such as previous technology disruptions, infrastructure failures, or health emergencies in the region and adjusting for AI-specific factors. It can be credibly done with only desk research and access to public databases, avoiding the need for expensive primary data collection.

Scenario planning complements this by enabling researchers to test whether policy responses would remain viable under different futures without requiring quantitative precision. The key adaptation we argue for is iterative implementation. Researchers should begin with lightweight assessments and deepen analysis only for risks that cross set severity thresholds. It can also be helpful to partner with regional institutions or join existing technical working groups such as those coordinated by Africa CDC to access shared frameworks and validation at minimal cost.

Depending on the resources available, researchers should also consider designing and using low-cost versions of SEE.

### 3.3.2 Poor internet penetration[135]

Low connectivity creates a distinctive threat landscape that requires scenario planning focused on tracing indirect AI exposure pathways rather than assuming direct user interaction. In environments where many people are offline, AI harms can manifest through intermediary channels. These include AI systems used by external actors making decisions about local populations without local visibility, or AI-generated disinformation spreading through non-digital channels such as radio, SMS, or in-person networks and reaching populations with high vulnerability to manipulation and no capacity to verify claims. In such contexts, AI amplifies latent risks of mass violence by enabling adversaries to produce incitement content at scale while the absence of digital literacy and robust fact-checking infrastructure removes barriers to sudden escalation.

Scenario planning's strength, which lies in testing whether policy responses remain viable under different conditions, applies well here because it can explore how risks propagate through these indirect pathways. Structured Expert Elicitation and participatory scenario planning workshops can also be adapted for offline implementation, using the kind of community-based methods described in section 3.2.3. The digital divide itself should be incorporated as a vulnerability factor within whatever method is employed.

### 3.3.3 Limited technical expertise[136]

Hybrid structured expert elicitation (SEE) becomes essential when technical expertise is thin, because it deliberately combines local contextual knowledge with remote technical input. As noted in section 3.2.2, SEE needs to be adapted to avoid over-reliance on Western technical experts who may not understand local contexts. The hybrid model enables AI technical expertise (potentially accessed through studying recent contributions or through supported consultations) to estimate model capabilities, while local infrastructure, health, and governance experts estimate what would happen if those capabilities were deployed in African settings. This division of labour compensates for the absence of individuals who possess both AI technical depth and local contextual understanding. Fact-finders can also leverage scenario planning through participatory workshops involving local communities, which section 3.2.3 identifies as a method of ensuring that mapped scenarios reflect lived realities. This approach mobilises non-technical forms of knowledge that remain highly relevant to threat assessment even without specialised AI expertise.

### 3.3.4 Weak state capacity[137]

Weak state capacity is especially relevant to severe AI risk because it increases the probability of harm materialising while limiting the mitigations available to address it. The absence of preventive structures increases the probability that adversaries can access and operationalise dangerous capabilities, while limited institutional coordination reduces the feasibility of implementing safeguards once risks are identified.

System-Theoretic Process Analysis (STPA) is particularly well-suited to weak state capacity contexts because it treats safety as a control problem rather than a reliability issue, and weak state capacity is precisely a breakdown in control structures. STPA examines how accidents occur when control loops between AI systems, human operators, sensors, and regulators fail, meaning it is ideal for identifying where governance gaps become primary risk vectors. As discussed in section 3.2.4, in Africa, STPA can model environmental instability as a key variable and thus be used to identify key weaknesses and build more fail-safe designs.

The critical adaptation is treating state capacity indicators including enforcement capability, regulatory coherence, and institutional coordination as explicit variables in the control loop analysis and incorporating "implementation probability" discounts that recognise identified risks may not trigger effective state responses.

### 3.3.5 Significant conflict[138]

Hybrid Threat Modelling (HTM) is the most appropriate for conflict contexts because it can model AI's interaction with social stability and combine multiple analytical techniques to identify threats within specific cultural contexts and feedback loops. HTM's combination of different methodologies and consideration of specific cultural context and processes enables it to capture how AI risks interact with existing conflict dynamics: group grievances, factional polarisation, external support for armed groups, and the legitimacy of the security apparatus.

Finally, context-aligned dangerous capability evaluations can be redesigned around conflict-specific risks. They can test for capabilities that could amplify latent conflict risks to mass casualty scale, for example by generating incitement content calibrated to local ethnic or political fault lines that could trigger sudden mass violence, providing tactical or targeting information that could enable atrocities, or assisting adversaries in acquiring biological or chemical materials. The likelihood of such risks is heightened in conflict zones due to the presence of numerous potential bad actors and the absence of functional preventive structures.

The preceding sections have focused on threat modelling for severe AI risks as informed by their context, that is, the process pathways through which risks could arise given the vulnerabilities, hazards, and exposure surfaces that characterise social-structural systems. Under this framing, frontier AI is assumed to serve primarily as an amplifier of latent risk rather than as the central risk vector. A separate question, however, is whether there could be Africa-centric threat models for misalignment risk, where frontier AI itself is the central source of danger.

# 4. Investigating Africa-centric models of AI misalignment risk

Whether there are distinct, exclusively Africa-centric threat models for misalignment depends on how misalignment is defined. Current artificial intelligence systems, specifically machine learning models, are trained by solving large-scale optimisation problems.[139,140] This process involves maximising or minimising a specific function relative to a set of constraints, often representing the range of choices available in a given situation. This function allows for the comparison of different choices to determine which output is "best." The potential for misalignment, where the model fails to adhere to human intent,[141] arises from how this maximisation is executed, how the function is constructed, or how the range of choices is defined. When these components are analysed individually, the argument for a unique African threat model dissolves into a case for universal applicability.

The first component to consider is the mechanism of optimisation itself. The mathematical and computational approaches used to update model weights, whether gradient-based or gradient-free, are universal. There is no evidence of a distinctly “African” method of mathematical optimisation; the underlying logic remains constant regardless of geography. Consequently, any misalignment arising from the optimisation process itself is not context dependent. This includes the phenomenon of reward hacking, where high optimisation pressure on a proxy reward function leads a model to exploit loopholes in the mathematical specification rather than achieving the intended goal.[142] Because the language of optimisation is universal, a model trained to optimise a specific metric will exploit that metric whether deployed in Lagos or London.

The second component is the function being optimised, which corresponds to the model or algorithm. Due to the democratisation of science and the global nature of technology, models developed or deployed in Africa utilise the same foundational architectures, such as Transformers or convolutional neural networks (CNNs), as those used globally. These architectures carry specific inductive biases, mathematical assumptions used to prioritise certain patterns over others.[143,144] Such biases can lead to goal misgeneralisation, where a model competently pursues an unintended objective because that objective was perfectly correlated with the intended goal during training. While subtle structural variations may exist, the mathematics governing these architectures ensures that the mapping of inputs to outputs follows the same fundamental rules. Structural vulnerabilities arising from model design are therefore not unique to the continent.

The third component, the definition of what constitutes the best output and the range of available choices, presents the most nuance, as it is derived from data distributions.[145] The set of cultural preferences, values, and environmental contexts in Africa differs from those in Western, Asian, and other datasets. Current research in interpretability demonstrates that data distributions heavily influence model behaviour,[146] and one might argue that if the African data distribution is substantially different, it could give rise to a unique threat model. However, this conclusion does not hold when examined through the lens of statistical and machine learning. Any specific real dataset, whether Western or African, is an empirical approximation drawn from a latent, “true” distribution of humanity.[147] Because humans share the same species-level needs, intellectual capabilities, and fundamental drivers, human actions exist within a universal range of possibilities. African data represents a subset of this general human distribution, intersecting with Western data in many areas while diverging in others. Distributional shift, the gap between the training distribution and the deployment environment, is therefore the mechanism that reveals latent misalignment. Any misalignment arising from these data differences is a universal problem of specification and generalisation, not a unique class of threat.

Finetuning and transfer learning do not, in themselves, create a new class of Africa-centric alignment mechanisms. Rather, they alter which points in the model's pre-existing hypothesis space are selected by changing the loss, the data distribution, and the optimisation trajectory. In practice, any behavioural change introduced by finetuning can be decomposed into familiar components: the objective being optimised, the structure and sampling density of the finetuning data, the inductive biases of the architecture, and the exposure conditions at deployment.

This does not imply that the African context is irrelevant to misalignment. Rather, it suggests that the continent acts as a distinct environment that may surface universal risks in particular ways. Misalignment scenarios such as reward hacking, goal misgeneralisation, or failures due to distributional shift can occur anywhere. However, specific features of the African context, such as low-resource linguistic environments or distinct socio-economic constraints, may trigger or expose these latent failures in ways that other contexts do not. Africa provides a specific set of conditions where this mismatch may be more consequential.

For instance, an AI model with biological reasoning capabilities poses a severe risk in regions with limited regulatory oversight. A misaligned model that fails to internalise safety constraints across all linguistic or technical jailbreak prompts could be used to design or synthesise toxic agents. In the African context, where public health infrastructure is already strained, the release of such an agent would represent a tail-risk event, amplified by the speed of AI-assisted design. These are not new threat models created by the African context, but existing universal threats revealed by it. Addressing these optimisation and alignment challenges in African contexts therefore implies addressing them for humanity at large, as the underlying mechanisms of failure are shared across all borders.

# 5. Practicalities of running severe AI risk evaluations

If the case for Africa-centric severe AI risk evaluations is accepted, the question becomes how to conduct them in practice. Such evaluations would not differ in their goals from other evaluations; they would still undertake a systematic and replicable process of eliciting model behaviour to measure safety-relevant properties under specified assumptions. What would likely shift are the constraints within which these evaluations are conducted. While these limitations do not undermine the relevance of existing evaluation frameworks, they do affect how those frameworks would be implemented in African contexts and, specifically, under what trade-offs.

Open-source and extensible frameworks are particularly valuable because they lower the barrier to designing, running, and iterating on end-to-end safety evaluations without requiring bespoke infrastructure. For example, Inspect,[148] built by the UK AI Security Institute, enables researchers to construct reproducible, task-based evaluations that make deployment assumptions explicit, allowing systematic testing under multilingual or low-resource conditions. Similarly, Control Arena[149]-style evaluations for agentic behaviour and autonomy can be run with tasks reflecting African deployment realities, such as planning under resource scarcity or other distributional shift cases, with little to no change to the underlying evaluation logic. The primary advantage of these tools is that they support systematic variation of prompts, environments, and monitoring assumptions.

These tools do, however, have limitations that matter for Africa-centric work. They do not, by default, solve issues of linguistic coverage or data availability, and their outputs remain sensitive to task design choices that fail to reflect the environmental volatility and socio-technical constraints of the continent. For instance, an evaluation that measures a model's helpfulness in a stable sandbox may fail to capture how that same model might misgeneralise its goals when faced with the sudden data sparsity, resource scarcity, or resource abundance characteristic of many African deployment environments. Meaningful use of these frameworks therefore requires not only local data, but a rethinking of the assumptions embedded in the test environment itself.

Beyond tool selection, the practical conditions under which evaluations are conducted also shape their validity.

To be useful, evaluations must reflect the conditions under which systems are actually deployed and used.[150,151] This includes realistic language inputs, patterns of code-switching, intermittent connectivity, the role of human oversight in decision-making, and integration with existing institutional processes. Designing evaluations that capture these realities is challenging, particularly when deployment environments differ substantially from those assumed by leading AI labs. Simplified or idealised evaluation setups risk overlooking failures that only surface under the conditions in which systems are actually used.

The cost of running evaluations presents a further challenge. Many safety evaluations require access to advanced AI models, significant computing power, or paid services, which can be prohibitively expensive for researchers and institutions in African contexts. As of early 2026, the AI landscape is divided between high-tier proprietary models and increasingly capable open-weight alternatives, each presenting different financial barriers.

Evaluating high-capability proprietary systems such as GPT-5.2, Gemini 3 Pro, or Grok 4.1, or using them as judges in evaluation pipelines, requires researchers to absorb significant per-token API fees. At a minimum of $10 per million output tokens, a research project requiring 50 million tokens[152] could reach $6,000 per month before accounting for the number of models evaluated, input tokens, and troubleshooting. This creates a significant "safety tax" for African institutions lacking Western-level endowments. Running open-weight models such as Llama 4 (405B+), DeepSeek-V3 (671B MoE), or Qwen 3 Max (235B) to avoid recurring fees requires steep hardware investment instead. Large frontier models necessitate data-centre-grade clusters costing between $40,000 and $300,000 or more on the refurbished market, while even mid-size 70B models require prosumer setups priced around $4,500 to $9,000. These costs increase further when evaluations must cover multiple languages, domains, or scenarios.

These resource constraints require careful prioritisation of which evaluations are most useful for reducing severe risks, favouring methods that can be adapted for different uses and identify multiple types of risks at once. Several strategies can help navigate these financial and technical barriers. Tiered evaluation pipelines that use smaller or lighter models to filter high-throughput tasks, escalating only complex reasoning cases to expensive frontier models, can reduce costs substantially. This can be furthered by leveraging recent quantisation techniques, such as 4-bit or 2-bit formats, to run frontier-class open models on cheaper consumer hardware, bypassing recurring API fees.

Beyond technical efficiencies, securing external support is equally important. Compute-for-research grants from frontier labs such as OpenAI and Google DeepMind, as well as programmes like NVIDIA Inception or AWS Cloud Credits for Research, can provide significant resources. Collaborative approaches also hold promise, including pooling demand for shared GPU clusters or utilising continent-focused compute-as-a-service providers to access data-centre-grade infrastructure at lower cost.

Finally, where local data is sparse, frontier models can, to some degree, be used to generate synthetic deployment scenarios reflective of African conditions, allowing researchers to test smaller local models without the prohibitive cost of live data collection.

Time relevance and data availability also constrain Africa-centric evaluations. In many domains of interest, such as public health, infrastructure, or governance, relevant local datasets may be sparse or non-existent, requiring new data collection, validation, and analysis. These processes are time-intensive and can reduce the timeliness of evaluation findings in rapidly evolving deployment contexts. A tension exists between contextual fidelity and responsiveness: highly localised evaluations may be more accurate but arrive too late to inform deployment decisions, while faster evaluations may rely on proxies that only partially capture local realities.

Together, these limitations point towards a pragmatic approach to evaluation. Rather than attempting to eliminate all uncertainty, Africa-centric evaluation efforts may benefit from methods that balance realism, cost, and timeliness, with attention to early detection of failures that could lead to critical harm under plausible worst-case conditions. Existing evaluation frameworks can be adapted to African contexts, provided their limitations and the uncertainties that persist are made explicit.

# 6. Recommendations

This paper has argued that Africa's deployment context creates safety gaps that current evaluation designs do not adequately capture, and that dedicated threat models and evaluation frameworks are needed to address them. From this analysis, several recommendations follow for the principal stakeholders involved in the development, evaluation, and overall governance of frontier AI systems.

**Frontier AI developers should revise their safety and preparedness frameworks to incorporate context-sensitive thresholds for economic harm.**
As this paper has shown, a uniform threshold of hundreds of billions of dollars in damage effectively renders the economic risks facing most African countries invisible to the evaluation process, since losses well below that figure could prove irrecoverable for smaller economies. Adopting relative measures, such as the 5% of GDP threshold proposed here, or developing comparable alternatives calibrated to the economic capacity of deployment regions, would produce a more accurate picture of where frontier AI poses severe risk.

**Frontier AI developers should expand the scope of their AI safety evaluations to include deployment conditions characteristic of the Global South.**
Because AI safety determinations shape deployment decisions, the environments assumed in evaluation design should reflect the environments in which these systems will operate, including in Africa. Where AI developers lack the contextual expertise to design such evaluations internally, partnerships with African research institutions and domain specialists offer a practical path forward. Additionally, frontier labs are well positioned to support the broader evaluation ecosystem by extending compute-for-research grants, API access programmes, and technical mentorship to African researchers conducting serious safety work.

These investments are a condition of safety, given that the robustness of safety claims depends on the breadth and representativeness of the evaluations underpinning them.

**African policymakers should take steps to ensure that evaluation requirements and safety standards account for the full range of deployment contexts.**

African governments that are actively developing national AI strategies, policies and laws should ensure that these instruments incorporate detailed safety evaluation requirements proportional to the risks involved. This is especially important given that the current policy momentum towards AI adoption on the continent is not matched by sufficient and comparable attention to the conditions under which deployed systems have been assessed for safety. Furthermore, policymakers in Africa should invest in building local AI evaluation capacity, including enacting legislation supporting the development of local datasets, funding African AI safety research institutions, and facilitating knowledge exchange between national AI safety bodies and their counterparts in developer jurisdictions.

**AI safety researchers and evaluators should adopt and build on the thinking and frameworks proposed in this paper when designing evaluations for Africa-centric severe AI risks.**

The taxonomy developed in this paper, comprising outcome thresholds, process pathways, and the distinctive features of amplification and time, is intended as a starting point rather than a final specification. In particular, the framework's focus on amplification of existing risks leaves a gap in accounting for genuinely novel risk categories, and further work is needed to address this limitation. Researchers should draw on the threat-modelling methodologies surveyed in this paper, selecting and combining techniques appropriate to the constraints of their operating environment, and should treat those constraints themselves as variables within their threat models.

# 7. Conclusion

This paper began with the observation that frontier AI systems are being deployed in contexts that differ materially from those in which they are developed and evaluated. It has sought to demonstrate that these differences run deep, that they are woven into the conditions under which people encounter these systems, and that they open pathways to severe harm that current evaluation designs are not configured to detect. The taxonomy, threat-modelling strategies, and practical guidance offered here are intended to support a necessary expansion of the evaluation ecosystem. They, of course, do not resolve the challenge, but they provide a foundation on which more representative and rigorous safety assessments can be built. If the interaction space that evaluations can cover will always remain a fraction of the whole, it matters a great deal which fraction is chosen. We hope to have made the case that Africa should be part of it.

[24] Department for Science, Innovation and Technology. (2025, February 7). *Frontier AI safety commitments: AI Seoul Summit 2024*. GOV.UK. https://www.gov.uk/government/publications/frontier-ai-safety-commitments-ai-seoul-summit-2024/frontier-ai-safety-commitments-ai-seoul-summit-2024

[25] Organisation for Economic Co-operation and Development (OECD). (2023). *Explanatory memorandum on the updated OECD definition of an AI system*. https://www.oecd.org/en/publications/explanatory-memorandum-on-the-updated-oecd-definition-of-an-ai-system_623da898-en.html

[26] Anthropic. (2023, September 19). *Responsible scaling policy* (Version 1.0). https://www-cdn.anthropic.com/1adf000c8f675958c2ee23805d91aaade1cd4613/responsible-scaling-policy.pdf

[27] OpenAI**.** (2025, April 15). *Preparedness framework* (Version 2). https://cdn.openai.com/pdf/18a02b5d-6b67-4cec-ab64-68cdfbddebcd/preparedness-framework-v2.pdf

[28] Dragan, A., King, H., & Dafoe, A**.** (2024, May 17). *Introducing the frontier safety framework*. Google DeepMind Blog. https://deepmind.google/blog/introducing-the-frontier-safety-framework/

[29] The risk of threat actors using critical capabilities of deployed or exfiltrated models to cause harm.

[30] Where general-purpose AI agents are potentially misaligned, thereby difficult to control and posing a risk of severe harm.

[31] ML R&D capabilities may reduce society's overall ability to manage AI risks, which may serve as a substantial cross-cutting risk factor to other pathways to severe harm.

[32] California Legislative Information. (2025). *Senate Bill No. 53: Transparency in Frontier Artificial Intelligence Act* (Cal. Stat. Ch. 138, 2025–2026 Reg. Sess.). https://leginfo.legislature.ca.gov/faces/billTextClient.xhtml?bill_id=202520260SB53

[33] New York RAISE Act, 12 https://www.nysenate.gov/legislation/bills/2025/A6453/amendment/B

[34] Hendrycks, D., Mazeiuka, M., & Woodside, T. (2023, October 9). *An overview of catastrophic AI risks. Center for AI Safety.* https://safe.ai/ai-risk

[35] Hendrycks, D. (2024). *Introduction to AI safety, ethics and society*. Taylor & Francis. https://www.aisafetybook.com

[36] Anthropic. (2025, May 14). *Responsible Scaling Policy* (Version 2.2). https://www-cdn.anthropic.com/872c653b2d0501d6ab44cf87f43e1dc4853e4d37.pdf

[37] OpenAI. (2025, April 15). *Preparedness framework* (Version 2). https://cdn.openai.com/pdf/18a02b5d-6b67-4cec-ab64-68cdfbddebcd/preparedness-framework-v2.pdf

[38] Google DeepMind. (2025, September 22). *Google Frontier safety framework* (Version 3.0). https://storage.googleapis.com/deepmind-media/DeepMind.com/Blog/strengthening-our-frontier-safety-framework/frontier-safety-framework_3.pdf

[39] Model Evaluation & Threat Research (METR). (n.d.). *Common elements of frontier AI safety policies*. https://metr.org/common-elements

[40] AI Security Institute. (n.d.). *Research agenda*. The AI Security Institute. https://www.aisi.gov.uk/research-agenda

[41] Apollo Research. (2025). *Security at Apollo Research*. https://www.apolloresearch.ai/blog/security-at-apollo-research/

[42] Apollo Research. (n.d.). *Apollo Research is becoming a PBC*. https://www.apolloresearch.ai/blog/apollo-research-is-becoming-a-pbc/

[43] Model Evaluation & Threat Research (METR). (n.d.). *About*. https://metr.org/about

[44] Grey, M., & Segerie, C.-R. (2025, June 13). *Safety by measurement: A systematic literature review of AI safety evaluation methods*. *arXiv*. https://arxiv.org/abs/2505.05541

[45] Ji, J., Venkatram, V., & Batalis, S. (2025, May 28). AI safety evaluations: An explainer. Center for Security and Emerging Technology. https://cset.georgetown.edu/article/ai-safety-evaluations-an-explainer/

[46] Grey, M., & Segerie, C.-R. (2025, June 13). *Safety by measurement: A systematic literature review of AI safety evaluation methods*. *arXiv*. https://arxiv.org/abs/2505.05541

[47] Grey, M., & Segerie, C.-R. (2025, June 13). *Safety by measurement: A systematic literature review of AI safety evaluation methods*. *arXiv*. https://arxiv.org/abs/2505.05541

[48] Grey, M., & Segerie, C.-R. (2025, June 13). *Safety by measurement: A systematic literature review of AI safety evaluation methods*. *arXiv*. https://arxiv.org/abs/2505.05541

[49] Greenblatt, R., Bucknall, B., Shlegeris, C., & Fabre, F. (2024, July 23). *AI control: Improving safety despite intentional subversion*. *arXiv*. https://arxiv.org/abs/2312.06942

[50] National Telecommunications and Information Administration. (2024, March 27). Purpose of evaluation. https://www.ntia.gov/issues/artificial-intelligence/ai-accountability-policy-report/developing-accountability-inputs-a-deeper-dive/ai-system-evaluations/purpose-of-evaluation

[51] Bommasani, R., Arora, S., Choi, Y., Fei-Fei, L., Ho, D. E., Jurafsky, D., Koyejo, S., Lakkaraju, H., Narayanan, A., Nelson, A., Pierson, E., Pineau, J., Varoquaux, G., Venkatasubramanian, S., Stoica, I., Liang, P., & Song, D. (n.d.). *A path for science- and evidence-based AI policy*. https://understanding-ai-safety.org/

[52] Artificial Intelligence Act, Regulation (EU) 2024/1689, art. 6, 2024 O.J. (L 1689). https://eur-lex.europa.eu/legal-content/EN/TXT/?uri=CELEX:32024R1689. Under the EU AI Act, an AI system is classified as high-risk where it is used as a safety component of a product subject to EU harmonisation legislation and mandatory third-party conformity assessment, or where it falls within the use cases listed in Annex III. Annex III systems may be exempted if they do not pose a significant risk to health, safety, or fundamental rights, except in cases involving profiling, which are always considered high-risk.

[53] Artificial Intelligence Act, Regulation (EU) 2024/1689, art. 9, 2024 O.J. (L 1689). https://eur-lex.europa.eu/legal-content/EN/TXT/?uri=CELEX:32024R1689

[54] Artificial Intelligence Act, Regulation (EU) 2024/1689, art. 72, 2024 O.J. (L 1689). https://eur-lex.europa.eu/legal-content/EN/TXT/?uri=CELEX:32024R1689. Article 72 requires providers of high-risk AI systems to establish and document a post-market monitoring system to document safety-relevant data.

[55] New York RAISE Act, 12 https://www.nysenate.gov/legislation/bills/2025/A6453/amendment/B

[56] California Legislative Information. (2025). Senate Bill No. 53: Transparency in Frontier Artificial Intelligence Act (Cal. Stat. Ch. 138, 2025–2026 Reg. Sess.). https://leginfo.legislature.ca.gov/faces/billTextClient.xhtml?bill_id=202520260SB53

[57] National Institute of Standards and Technology, Artificial Intelligence Risk Management Framework (AI RMF 1.0), January 2023, pg. 11. https://share.google/qgT4gY7sPSibykjS3.

[58] Liao, T. I., Taori, R., Raji, I. D., & Schmidt, L. (2021). Are we learning yet? A meta-review of evaluation failures across machine learning. Defined external validity as "the ability to make valid conclusions for contexts outside the experimental parameters" (p. 3); Shwartz, S., [Other Authors' Initials]., & [Last Author's Initial]. (2025, May 30). *Reality check: A new evaluation ecosystem is necessary to understand AI's real work effects*. *arXiv*. https://arxiv.org/html/2505.18893v4; Keller D, Steed R, Bergman S (2025, December 2). Accelerating AI innovation through measurement science. *National Institute of Standards and Technology*. https://www.nist.gov/blogs/caisi-research-blog/accelerating-ai-innovation-through-measurement-science.
[59] Selbst, A. D., Boyd, D., Friedler, S. A., Venkatasubramanian, S., & Vertesi, J. (2019, January). Fairness and abstraction in sociotechnical systems. *Proceedings of the 2019 Conference on Fairness, Accountability, and Transparency*, 59–68. https://doi.org/10.1145/3287560.3287598
[60] Selbst, A. D., Boyd, D., Friedler, S. A., Venkatasubramanian, S., & Vertesi, J. (2019, January). Fairness and abstraction in sociotechnical systems. *Proceedings of the 2019 Conference on Fairness, Accountability, and Transparency*, 59–68. https://doi.org/10.1145/3287560.3287598
[61] Grey, M., & Segerie, C.-R. (2025, August 26). *Safety by measurement: A systematic literature review of AI safety evaluations methods*. AI Safety Atlas, p.108-110. https://ai-safety-atlas.com/chapters/05
[62] Grey, M., & Segerie, C.-R. (2025, August 26). *Safety by measurement: A systematic literature review of AI safety evaluations methods*. AI Safety Atlas, p.108-110. https://ai-safety-atlas.com/chapters/05
[63] Grey, M., & Segerie, C.-R. (2025, August 26). *Safety by measurement: A systematic literature review of AI safety evaluations methods*. AI Safety Atlas. https://ai-safety-atlas.com/chapters/05
[64] https://au.int/en/documents/20240809/continental-artificial-intelligence-strategy
[65] African Union. (2024). *Continental artificial intelligence strategy: Harnessing AI for Africa's development and prosperity*, p. 4. https://au.int/sites/default/files/documents/44004-doc-EN-_Continental_AI_Strategy_July_2024.pdf
[66] For example, Benin https://numerique.gouv.bj/assets/documents/national-artificial-intelligence-and-big-data-strategy-1682673348.pdf; Egypt https://mcit.gov.eg/en/Publication/Publication_Summary/9283; Mauritius https://treasury.govmu.org/Documents/Strategies/Mauritius%20AI%20Strategy.pdf; Rwanda https://www.ictworks.org/wp-content/uploads/2023/12/Rwanda_Artificial_Intelligence_Policy.pdf;
[67] OpenAI. (2026, January 20). Horizon 1000: Advancing AI for primary healthcare. https://openai.com/index/horizon-1000/; Rigby, J. (2026, January 21). Gates and OpenAI team up for AI health push in African countries. Reuters. https://www.reuters.com/business/healthcare-pharmaceuticals/gates-openai-team-up-ai-health-push-african-countries-2026-01-21/
[68] Tshimula, J. M., Kalengayi, M., Makenga, D., Lilonge, D., Asumani, M., Madiya, D., Kalonji, É. N., Kanda, H., Galekwa, R. M., Kumbu, J., Mikese, H., Tshimula, G., Muabila, J. T., Mayemba, C. N., Nkashama, D. K., Kalala, K., Ataky, S., Basele, T. W., Didier, M. M., Kasereka, S. K., Dialufuma, M. V., Wa Kumwita, G. I., Muyuku, L., Kimpesa, J.-P., Muteba, D., Abedi, A. A., Ntobo, L. M., Bundutidi, G. M., Mashinda, D. K., Mpinga, E. K., & Kasoro, N. M. (2024, August 5). *Artificial intelligence for public health surveillance in Africa: Applications and opportunities* (arXiv:2408.02575). arXiv. https://doi.org/10.48550/arXiv.2408.02575
[69] IBM. (n.d.). What is critical infrastructure? https://www.ibm.com/think/topics/critical-infrastructure

https://ccianet.org/articles/an-examination-of-california-and-new-york-policy-on-regulating-frontier-ai/

[105] Liu, H., Lauta, K. C., & Maas, M. M. (2018). Governing boring apocalypses: A new typology of existential vulnerabilities and exposures for existential risk research. *Futures*, *102*, 6–19. https://doi.org/10.1016/j.futures.2018.04.009

[106] Liu, H., Lauta, K. C., & Maas, M. M. (2018). Governing boring apocalypses: A new typology of existential vulnerabilities and exposures for existential risk research. *Futures*, *102*, 6–19. https://doi.org/10.1016/j.futures.2018.04.009

[107] Liu, H., Lauta, K. C., & Maas, M. M. (2018). Governing boring apocalypses: A new typology of existential vulnerabilities and exposures for existential risk research. *Futures*, *102*, 6–19. https://doi.org/10.1016/j.futures.2018.04.009

[108] Our World in Data. (2024). *Annual number of deaths by cause, Africa, 2021*. https://ourworldindata.org/grapher/annual-number-of-deaths-by-cause?country=~OWID_AFR&tableFilter=continents

[109] Our World in Data. (2024). *Causes of death, 2021*. https://ourworldindata.org/grapher/annual-number-of-deaths-by-cause?tab=table&country=~OWID_EUR&tableFilter=continents

[110] Roser, M. (2022). *Malaria: One of the leading causes of child deaths, but progress is possible and you can contribute to it*. Our World in Data. https://ourworldindata.org/malaria-introduction

[111] Nájera, J. A., González-Silva, M., & Alonso, P. L. (2011). Some lessons for the future from the Global Malaria Eradication Programme (1955–1969). *PLoS Medicine*, *8*(1), Article e1000412. https://doi.org/10.1371/journal.pmed.1000412

[112] Nájera, J. A., González-Silva, M., & Alonso, P. L. (2011). Some lessons for the future from the Global Malaria Eradication Programme (1955–1969). *PLoS Medicine*, *8*(1), Article e1000412. https://doi.org/10.1371/journal.pmed.1000412

[113] Arnscheidt, C. W., (2025). Systemic contributions to global catastrophic risk. *Global Sustainability*, *8*, Article e19. https://doi.org/10.1017/sus.2025.20

[114] Stefánsson, H. O. (2020). Catastrophic risk. *Philosophy Compass*, *15*(11), Article e12709. https://doi.org/10.1111/phc3.12709 (quoting Posner).

[115] Grey, M., & Segerie, C.-R. (2025, June 13). Safety by measurement: A systematic literature review of AI safety evaluation methods. arXiv. https://arxiv.org/abs/2505.05541

[116] Larson, E. V. (2019). Force planning scenarios, 1945–2016: Their origins and use in defense strategic planning (Report No. RR-2173/1-A). RAND Corporation. https://doi.org/10.7249/RR2173.1

[117] Bodeau, D. J., McCollum, C. D., & Fox, D. B. (2018). Cyber threat modeling: Survey, assessment, and representative framework (Report No. 18-1174). MITRE Corporation. https://www.mitre.org/sites/default/files/2021-11/prs-18-1174-ngci-cyber-threat-modeling.pdf

[118] Cantarelli, C. C., Davis, K., Pinto, J. K., & Turner, N. (2025). Reference class forecasting: Promises, problems, and a research agenda moving forward. *Production Planning & Control*, p.2. https://doi.org/10.1080/09537287.2025.2578708

[119] Cantarelli, C. C., Davis, K., Pinto, J. K., & Turner, N. (2025). Reference class forecasting: Promises, problems, and a research agenda moving forward. *Production Planning & Control*, p.2. https://doi.org/10.1080/09537287.2025.2578708

[150] Segerie, C.-R., & Salle, J. (n.d.). Evaluation design. In AI Safety Atlas. https://ai-safety-atlas.com/chapters/v1/evaluations/evaluation-design

[151] Frontier Model Forum. (2025, April 22). Frontier capability assessments: Technical report. https://www.frontiermodelforum.org/uploads/2025/04/FMF-PDF-Frontier-Capability-Assessments_-Technical-Report.pdf

[152] Artificial Analysis. (n.d.). AI model & API providers analysis. Retrieved February 13, 2026, from https://artificialanalysis.ai/